\RequirePackage{lineno}
\documentclass[aps,prl,showpacs,twocolumn,superscriptaddress,letterpaper]{revtex4}

\usepackage{graphicx}
\usepackage{dcolumn}
\usepackage{bm}
\usepackage{amsmath, amssymb}%
\usepackage{epstopdf}
\usepackage{color}

\usepackage{hhline}
\usepackage{multirow}

\usepackage{float}
\floatstyle{plaintop}
\restylefloat{table}

\newcommand{\eepp}{\mbox{($e,e'pp$)}}
\newcommand{\eep}{\mbox{($e,e'p$)}}

\newcommand{\He}{$^{4}$He}
\newcommand{\pcm}{$\vec p_{c.m.}$}
\newcommand{\prel}{$\vec p_{rel}$}

\usepackage[]{changes} 

\begin{document}

\title{The center of mass motion of short-range correlated nucleon pairs studied via the $A$\eepp{} reaction}

\newcommand*{\TAU }{School of Physics and Astronomy, Tel Aviv University, Tel Aviv 69978, Israel}
\newcommand*{\TAUindex}{1}
\affiliation{\TAU} 
\newcommand*{\MIT }{Massachusetts Institute of Technology, Cambridge, Massachusetts 02139, USA}
\newcommand*{\MITindex}{2}
\affiliation{\MIT} 
\newcommand*{\ODU}{Old Dominion University, Norfolk, Virginia 23529}
\newcommand*{\ODUindex}{3}
\affiliation{\ODU} 
\newcommand*{\UTFSM}{Universidad T\'{e}cnica Federico Santa Mar\'{i}a, Casilla 110-V Valpara\'{i}so, Chile}
\newcommand*{\UTFSMindex}{4}
\affiliation{\UTFSM}
\newcommand*{\ANL}{Argonne National Laboratory, Argonne, Illinois 60439}
\newcommand*{\ANLindex}{5}
\affiliation{\ANL}
\newcommand*{\ASU}{Arizona State University, Tempe, Arizona 85287-1504}
\newcommand*{\ASUindex}{6}
\affiliation{\ASU}
\newcommand*{\CSUDH}{California State University, Dominguez Hills, Carson, CA 90747}
\newcommand*{\CSUDHindex}{7}
\affiliation{\CSUDH}
\newcommand*{\CANISIUS}{Canisius College, Buffalo, NY}
\newcommand*{\CANISIUSindex}{8}
\affiliation{\CANISIUS}
\newcommand*{\CMU}{Carnegie Mellon University, Pittsburgh, Pennsylvania 15213}
\newcommand*{\CMUindex}{9}
\affiliation{\CMU}
\newcommand*{\CUA}{Catholic University of America, Washington, D.C. 20064}
\newcommand*{\CUAindex}{10}
\affiliation{\CUA}
\newcommand*{\SACLAY}{IRFU, CEA, Universit'e Paris-Saclay, F-91191 Gif-sur-Yvette, France}
\newcommand*{\SACLAYindex}{11}
\affiliation{\SACLAY}
\newcommand*{\CNU}{Christopher Newport University, Newport News, Virginia 23606}
\newcommand*{\CNUindex}{12}
\affiliation{\CNU}
\newcommand*{\UCONN}{University of Connecticut, Storrs, Connecticut 06269}
\newcommand*{\UCONNindex}{13}
\affiliation{\UCONN}
\newcommand*{\DUKE}{Duke University, Durham, North Carolina 27708-0305}
\newcommand*{\DUKEindex}{14}
\affiliation{\DUKE}
\newcommand*{\FU}{Fairfield University, Fairfield CT 06824}
\newcommand*{\FUindex}{15}
\affiliation{\FU}
\newcommand*{\FERRARAU}{Universita' di Ferrara , 44121 Ferrara, Italy}
\newcommand*{\FERRARAUindex}{16}
\affiliation{\FERRARAU}
\newcommand*{\FIU}{Florida International University, Miami, Florida 33199}
\newcommand*{\FIUindex}{17}
\affiliation{\FIU}
\newcommand*{\FSU}{Florida State University, Tallahassee, Florida 32306}
\newcommand*{\FSUindex}{18}
\affiliation{\FSU}
\newcommand*{\INFNFE}{INFN, Sezione di Ferrara, 44100 Ferrara, Italy}
\newcommand*{\INFNFEindex}{19}
\affiliation{\INFNFE}
\newcommand*{\INFNFR}{INFN, Laboratori Nazionali di Frascati, 00044 Frascati, Italy}
\newcommand*{\INFNFRindex}{20}
\affiliation{\INFNFR}
\newcommand*{\INFNGE}{INFN, Sezione di Genova, 16146 Genova, Italy}
\newcommand*{\INFNGEindex}{21}
\affiliation{\INFNGE}
\newcommand*{\INFNRO}{INFN, Sezione di Roma Tor Vergata, 00133 Rome, Italy}
\newcommand*{\INFNROindex}{22}
\affiliation{\INFNRO}
\newcommand*{\INFNTUR}{INFN, Sezione di Torino, 10125 Torino, Italy}
\newcommand*{\INFNTURindex}{23}
\affiliation{\INFNTUR}
\newcommand*{\ORSAY}{Institut de Physique Nucl\'eaire, CNRS/IN2P3 and Universit\'e Paris Sud, Orsay, France}
\newcommand*{\ORSAYindex}{24}
\affiliation{\ORSAY}
\newcommand*{\ITEP}{Institute of Theoretical and Experimental Physics, Moscow, 117259, Russia}
\newcommand*{\ITEPindex}{25}
\affiliation{\ITEP}
\newcommand*{\JMU}{James Madison University, Harrisonburg, Virginia 22807}
\newcommand*{\JMUindex}{26}
\affiliation{\JMU}
\newcommand*{\KNU}{Kyungpook National University, Daegu 41566, Republic of Korea}
\newcommand*{\KNUindex}{27}
\affiliation{\KNU}
\newcommand*{\MISS}{Mississippi State University, Mississippi State, MS 39762-5167}
\newcommand*{\MISSindex}{28}
\affiliation{\MISS}
\newcommand*{\UNH}{University of New Hampshire, Durham, New Hampshire 03824-3568}
\newcommand*{\UNHindex}{29}
\affiliation{\UNH}
\newcommand*{\NRCN}{Nuclear Research Centre Negev, Beer-Sheva, Israel}
\newcommand*{\NRCNindex}{30}
\affiliation{\NRCN}
\newcommand*{\NSU}{Norfolk State University, Norfolk, Virginia 23504}
\newcommand*{\NSUindex}{31}
\affiliation{\NSU}
\newcommand*{\OHIOU}{Ohio University, Athens, Ohio  45701}
\newcommand*{\OHIOUindex}{32}
\affiliation{\OHIOU}
\newcommand*{\RPI}{Rensselaer Polytechnic Institute, Troy, New York 12180-3590}
\newcommand*{\RPIindex}{33}
\affiliation{\RPI}
\newcommand*{\ROMAII}{Universita' di Roma Tor Vergata, 00133 Rome Italy}
\newcommand*{\ROMAIIindex}{34}
\affiliation{\ROMAII}
\newcommand*{\MSU}{Skobeltsyn Institute of Nuclear Physics, Lomonosov Moscow State University, 119234 Moscow, Russia}
\newcommand*{\MSUindex}{35}
\affiliation{\MSU}
\newcommand*{\SCAROLINA}{University of South Carolina, Columbia, South Carolina 29208}
\newcommand*{\SCAROLINAindex}{36}
\affiliation{\SCAROLINA}
\newcommand*{\TEMPLE}{Temple University,  Philadelphia, PA 19122 }
\newcommand*{\TEMPLEindex}{37}
\affiliation{\TEMPLE}
\newcommand*{\JLAB}{Thomas Jefferson National Accelerator Facility, Newport News, Virginia 23606}
\newcommand*{\JLABindex}{38}
\affiliation{\JLAB}
\newcommand*{\EDINBURGH}{Edinburgh University, Edinburgh EH9 3JZ, United Kingdom}
\newcommand*{\EDINBURGHindex}{39}
\affiliation{\EDINBURGH}
\newcommand*{\GLASGOW}{University of Glasgow, Glasgow G12 8QQ, United Kingdom}
\newcommand*{\GLASGOWindex}{40}
\affiliation{\GLASGOW}
\newcommand*{\VT}{Virginia Tech, Blacksburg, Virginia   24061-0435}
\newcommand*{\VTindex}{41}
\affiliation{\VT}
\newcommand*{\VIRGINIA}{University of Virginia, Charlottesville, Virginia 22901}
\newcommand*{\VIRGINIAindex}{42}
\affiliation{\VIRGINIA}
\newcommand*{\WM}{College of William and Mary, Williamsburg, Virginia 23187-8795}
\newcommand*{\WMindex}{43}
\affiliation{\WM}
\newcommand*{\YEREVAN}{Yerevan Physics Institute, 375036 Yerevan, Armenia}
\newcommand*{\YEREVANindex}{44}
\affiliation{\YEREVAN}
\newcommand*{\NOWISU}{Idaho State University, Pocatello, Idaho 83209}
\newcommand*{\GWU}{Institute for Nuclear Studies, Department of Physics, The George Washington University, Washington DC 20052, USA}
\newcommand*{\GWUindex}{45}
\affiliation{\GWU}

\author{E. O. Cohen}
\affiliation{\TAU}
\author{O.~Hen}
\email[Contact Author \ ]{hen@mit.edu}
\affiliation{\MIT}
\author{E.~Piasetzky}
\affiliation{\TAU}
\author{L.B.~Weinstein}
\affiliation{\ODU}
\author{M.~Duer}
\affiliation{\TAU}
\author{A.~Schmidt}
\affiliation{\MIT}
\author{I.~Korover}
\affiliation{\TAU}
\author {H.~Hakobyan} 
\affiliation{\UTFSM}
\author {S. Adhikari} 
\affiliation{\FIU}
\author {Z.~Akbar} 
\affiliation{\FSU}
% ...

\collaboration{The CLAS Collaboration}
\noaffiliation

\begin{abstract}

  Short-Range Correlated (SRC) nucleon pairs are a vital part of the
  nucleus, accounting for almost all nucleons with momentum greater
  than the Fermi momentum ($k_F$). A fundamental characteristic of SRC
  pairs is having large relative momenta as compared to $k_F$,
  and smaller center-of-mass (c.m.) which indicates a small separation
  distance between the nucleons in the pair. Determining the
  c.m. momentum distribution of SRC pairs is essential for
  understanding their formation process. We report here on the
  extraction of the c.m. motion of proton-proton ($pp$) SRC pairs in
  Carbon and, for the first time in heavier and ansymetric nuclei: aluminum, iron, and lead, from
  measurements of the $A$\eepp{} reaction. We find that the pair
  c.m. motion for these nuclei can be described by a three-dimensional
  Gaussian with a narrow width ranging from $140$ to $170$ MeV/$c$,
  approximately consistent with the sum of two mean-field nucleon
  momenta.  Comparison with calculations appears to show that the SRC
  pairs are formed from mean-field nucleons in specific quantum
  states. 

\end{abstract}

\maketitle

The atomic nucleus is a complex, strongly interacting, many body system. Effective theories can successfully describe the long-range part of the nuclear many-body wave function. However, the exact description of its short-range part is challenging. This difficulty is due to the complexity of the nucleon-nucleon ($NN$) interaction and the large nuclear density, which make it difficult to simplify the problem using scale-separated approaches when describing the short-range part of the nuclear wave-function.

Recent experimental studies have shown that approximately 20\% of the nucleons in the nucleus belong to strongly interacting, momentary, short-range correlated (SRC) nucleon pairs~\cite{frankfurt93,egiyan02,egiyan06,fomin12}. These pairs are predominantly proton-neutron pairs with a center-of-mass (c.m.) momentum $p_{c.m.}$ that is comparable to any two nucleons in the nuclear ground state and a much higher relative momentum $p_{rel}$ between the nucleons in the pair ($>k_F$, the nuclear Fermi momentum)~\cite{tang03,piasetzky06,shneor07,subedi08,korover14,hen14}. They account for almost all of the nucleons in the nucleus with momentum greater than $k_F$ and for $50\%$ to $60\%$ of the kinetic energy carried by nucleons in the nucleus~\cite{Vanhalst:2014cqa,wiringa14,Lonardoni:2017egu,CiofidegliAtti:1995qe,hen14}. See Refs.~\cite{Hen:2016kwk,Atti:2015eda,frankfurt08b} for recent reviews. SRC pairs are thus a vital part of nuclei with implications for many important topics including the possible modification of bound nucleon structure and the extraction of the free neutron structure function~\cite{Hen:2016kwk,weinstein11,Hen12,Hen:2013oha,hen11,Chen:2016bde}, neutrino-nucleus interactions and neutrino oscillation experiments~\cite{Gallagher:2011zza,fields13,Fiorentini13b,acciardi14,Weinstein:2016inx,VanCuyck:2016fab}, neutrino-less double beta decay searches~\cite{Kortelainen2007128,Song:2017ktj}, as well as neutron star structure and the nuclear symmetry energy~\cite{hen15,Cai:2015xga,Hen:2016ysx}.

The smaller c.m. momentum as compared to the large relative momentum of SRC pairs is a fundamental characteristic of such pairs, and is an essential indications that the nucleons in the pair are in close proximity with limited interaction with the surrounding nuclear environment~\cite{Weiss:2016obx}. 

Modern calculations \cite{Colle:2013nna} indicate that SRC pairs are temporary fluctuations due to the short-range part of the $NN$ interaction acting on two nucleons occupying shell-model (``mean-field'') states. The exact parentage and formation process of SRC pairs is not well understood. While state-of-the-art many-body calculations of one- and two-body momentum densities in nuclei~\cite{wiringa14,Carlson:2014vla,neff15} seem to produce SRC features that are generally consistent with measurements, they do not offer direct insight into the effective mechanisms of SRC pair formation.

Effective calculations using scale-separated approaches agree with many-body calculations ~\cite{Weiss:2016obx,Vanhalst:2014cqa,Alvioli:2016wwp,CiofidegliAtti:2017tnm}, suggesting that, at high-momenta, the momentum distribution of SRC pairs can be factorized into the c.m. and relative momentum distributions , 
\begin{equation}
n_{SRC}(\vec p_1, \vec p_2)\approx n^A_{c.m.}(\vec p_{c.m.}) n^{NN}_{rel}(\vec p_{rel}),
\label{eq:factorize}
\end{equation}
where $|\vec p_{rel}|$ is greater than $k_F$  and $|\vec p_{c.m.}| < |\vec p_{rel}|$~\cite{Weiss:2016obx, CiofidegliAtti:2017hhn, CiofidegliAtti:2017tnm}.
This implies that the relative momentum distribution of SRC pairs, $n^{NN}_{rel}(\vec p_{rel})$, is a universal function of the short-range part of the ($NN$) interaction, such that the many-body nuclear dynamics affect only the c.m. momentum distribution, $n^A_{c.m.}(\vec p_{c.m.})$. Therefore, extracting the c.m. momentum distribution of SRC pairs can provide valuable insight into their formation process.

The c.m. momentum distributions of SRC pairs in
\He\ and C have been extracted previously from $A(e,e'pN)$ and $A(p,2pn)$
measurements \cite{tang03,shneor07,korover14}. Here we present the
first study of the c.m. momentum distribution of $pp$ SRC pairs in
nuclei heavier than C using the $A$\eepp{} reaction.  
The cross-section for this \eepp{} two-nucleon knockout reaction 
in some kinematics approximately factorizes as a kinematic term times the elementary 
 electron-proton cross section times the nuclear decay
function, which defines the combined probability of finding the
knocked-out nucleon pair with given energies and
momenta~\cite{Frankfurt81,Frankfurt88,piasetzky06,Ryckebusch:1996wc,Colle:2013nna}.  
The decay function also factorizes into  relative and c.m. parts, just like
Eq. \ref{eq:factorize}~\cite{piasetzky06}.  Therefore, the  $A$\eepp{} cross section is 
approximately proportional to the c.m. momentum distributions of SRC
pairs~\cite{cda91,piasetzky06,sargsian05,Ryckebusch:1996wc,Colle:2013nna}:
\begin{equation}
\sigma\eepp \propto n^A_{c.m.}(\vec p_{c.m.}) .
\label{eq:sigmafactorize}
\end{equation}

To increase sensitivity to the initial state properties of $pp$-SRC
pairs, the measurement was done using high energy electrons scattering
at large momentum transfer (hard scattering), in kinematics dominated by the
hard breakup of SRC pairs, as discussed in detail
in~\cite{Hen:2016kwk}.
In this kinematics, Eq.~\ref{eq:sigmafactorize} is a good
approximation since rescattering of the two outgoing nucleons does not
distort the width of the momentum distribution (see discussion below).



\begin{figure} [t]
\includegraphics[width=6.5cm, height=5cm]{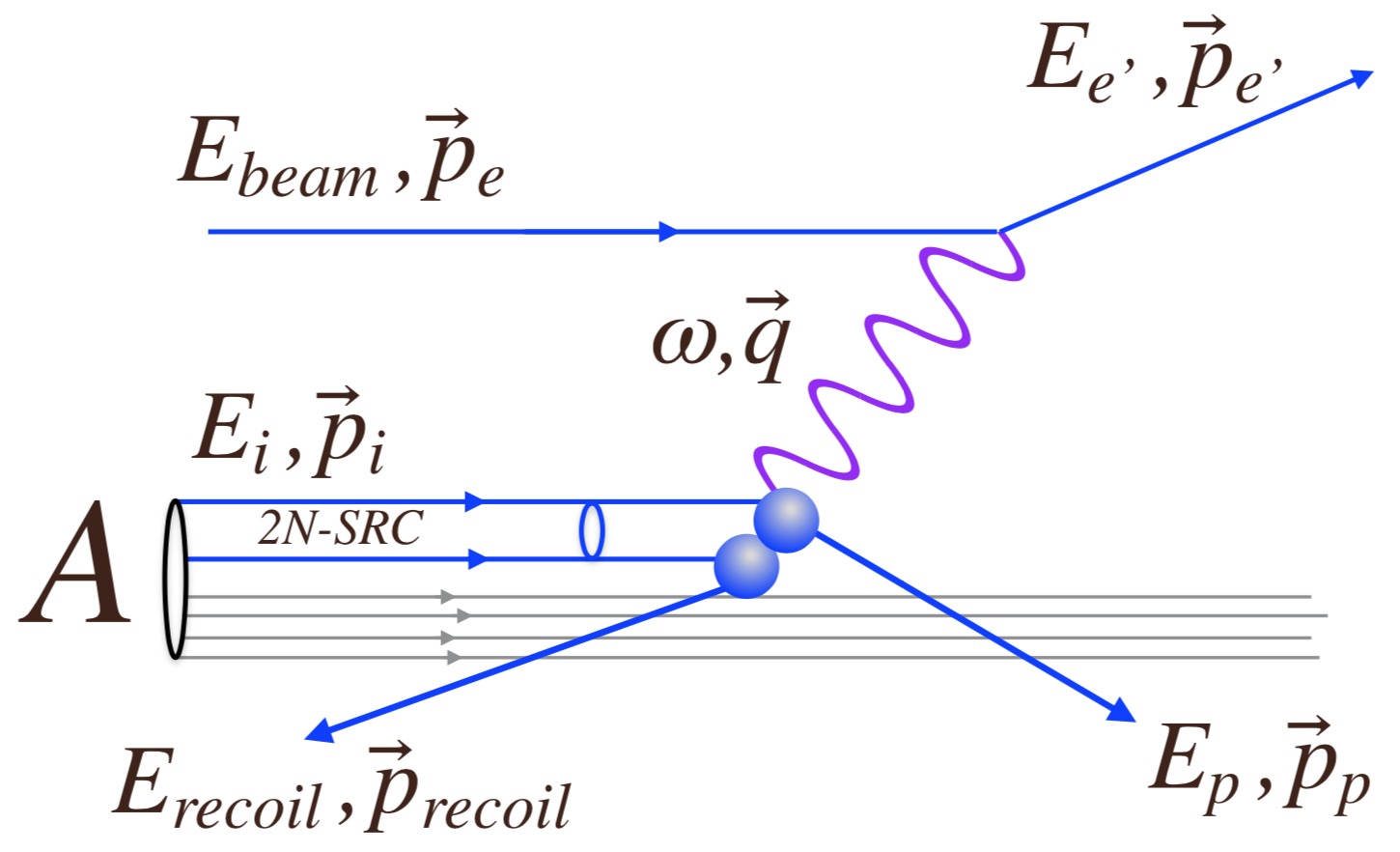}
\caption{\label{fig:0} (color online) Kinematics of the hard breakup of a $pp$-SRC pair in a hard two-nucleons knockout $A$\eepp{} reaction. See text for details.}
\end{figure}

The data presented here were collected as part of the EG2 run period that took place in 2004 in Hall B of the Thomas Jefferson National Accelerator Facility (Jefferson Lab). The experiment used a 5.01 GeV electron beam, impinging on $^2$H and natural C, Al, Fe, and Pb targets at the CEBAF Large Acceptance Spectrometer (CLAS)~\cite{Mecking:2003zu}. The analysis was carried out as part of the Jefferson Lab Hall B Data-Mining project.

CLAS used a toroidal magnetic field and six independent sets of drift chambers for charged particle tracking, time-of-flight scintillation counters for hadron identification, and {\v C}erenkov counters and electro-magnetic calorimeters for electron/pion separation. The polar angular acceptance was $8^o \le \theta \le140^o$ and the azimuthal angular acceptance ranged from 50\% at small polar angles to 80\% at larger polar angles. See Refs.~\cite{hen14,hen12a} for details on the electron and proton identification and momentum reconstruction procedures.

The EG2 run period used a specially designed target setup, consisting of an approximately 2-cm LD$_2$ cryotarget followed by one of six independently-insertable solid targets (thin Al, thick Al, Sn, C, Fe, and Pb, all with natural isotopic abundance, ranging between 0.16 and 0.38~g/cm$^2$), see Ref.~\cite{Hakobyan:2008zz} for details. The LD$_2$ target cell and the inserted solid target were separated by about 4 cm. The few-mm vertex reconstruction resolution of CLAS for both electrons and protons was sufficient to unambiguously separate particles originating in the cryotarget and the solid target. 

The kinematics of the  $A$\eepp{} reaction is shown schematically in Fig.~\ref{fig:0}. Identification of exclusive $A$\eepp{} events, dominated by scattering off $2N$-SRC pairs, was done in two stages: (1) selection of $A$\eep{} events in which the electron predominantly interacts with a single proton belonging to an SRC pair in the nucleus~\cite{subedi08,hen12a,hen14}, and (2) selection of $A$\eepp{} events by requiring the detection of a second, recoil, proton in coincidence with the $A$\eep{} reaction. 

We selected $A$\eep{} events in which the knocked-out proton predominantly belonged to an SRC pair by requiring a large Bjorken scaling parameter $x_B = Q^2 / (2m_p\omega) \geq 1.2$ (where $Q^2 = \vec q\thinspace^2 -\omega^2$, $\vec{q}$ and $\omega$ are the three-momentum and energy, respectively, transferred to the nucleus, and $m_p$ is the proton mass). This requirement also suppressed the effect of inelastic reaction mechanisms (e.g., pion and resonance production) and resulted in $Q^2 \geq 1.4$ GeV$^2$~\cite{Frankfurt:1996xx,shneor07}. We also required large missing momentum $300 \leq \vert\vec{p}_{miss}\vert \leq 600$ MeV/$c$, where $\vec{p}_{miss} = \vec{p}_{p} - \vec{q}$ with $\vec{p}_{p}$ the measured proton momentum. We further suppressed contributions from inelastic excitations of the struck nucleon by limiting the reconstructed missing mass of the two-nucleon system $m_{miss} = [(\omega+2m-E_p)^2 - p_{miss}^2]^{1/2} \leq 1.1$ GeV/$c^2$ (where $E_p$ is the total energy of the leading proton). We identified events where the leading proton absorbed the transferred momentum by requiring that its momentum $\vec{p}_{p}$ was within 25$^o$ of $\vec{q}$ and that $0.60 \leq |\vec{p}_{p}|/|\vec{q}| \leq 0.96$~\cite{hen12a,hen14}. As shown by previous experimental and theoretical studies, these conditions enhance the contribution of scattering off nucleons in SRC pairs and suppress contribution from competing effects \cite{groep00,Blomqvist:1998gq,Kester:1995zz,Arnold:1989qr,Laget:1987hb,Frankfurt:1996xx,Colle:2015lyl,Sargsian:2001ax}.

\begin{figure} [t]
\includegraphics[width=8cm, height=8.8cm]{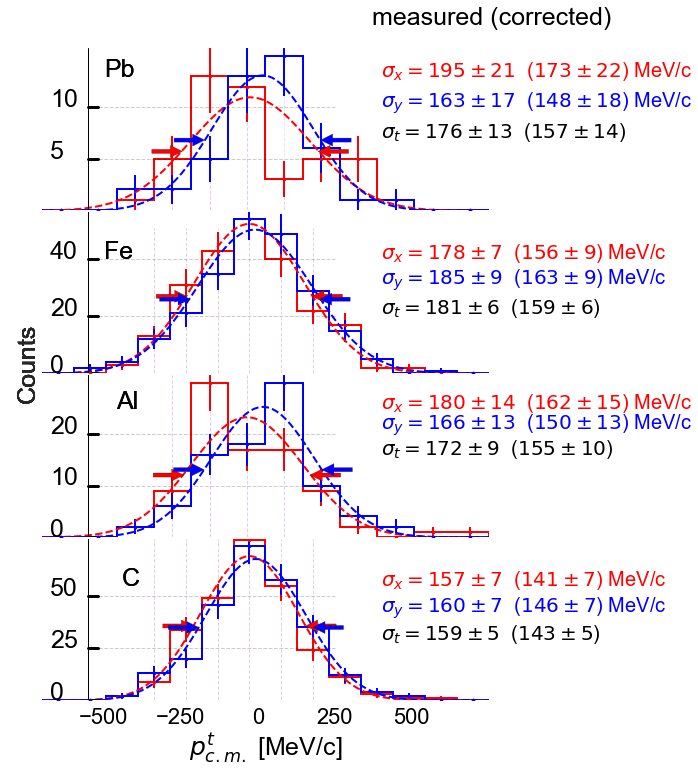}
\caption{\label{fig:1} (color online) The number of $A(e,e'pp)$ events plotted versus the  components of $\vec{p}_{c.m.}$ perpendicular to $\vec p_{miss}$. The red and blue histograms show the $\hat x$ and $\hat y$ directions, respectively. The data are shown before corrections for the CLAS detector acceptance. The dashed lines show the results of Gaussian fits to the data. The  widths in parentheses with uncertainties are corrected for the CLAS acceptance as discussed in the text.}
\end{figure}

$A$\eepp{} events were selected by requiring that the $A$\eep{} event had a second, recoil proton with momentum $|\vec{p}_{recoil}| \geq 350$ MeV/$c$. There were no events in which the recoil proton passed the leading proton selection cuts described above. The recoil proton was emitted opposite to $\vec{p}_{miss}$~\cite{hen14}, consistent with the measured pairs having large relative momentum and smaller c.m. momentum.

In the Plane Wave Impulse Approximation (PWIA), where the nucleons do not rescatter as they leave the nucleus, $\vec p_{miss}$ and $\vec p_{recoil}$ are equal to the initial momenta of the two protons in the nucleus before the interaction. In that case we can write
\begin{align}
\vec{p}_{c.m.} &= \vec{p}_{miss} + \vec{p}_{recoil} = \vec{p}_{p} - \vec{q} + \vec{p}_{recoil} \label{eq:2} \\
\vec p_{rel} &= \frac12(\vec{p}_{miss} - \vec{p}_{recoil}).
\end{align}
We use a coordinate system where $\hat z$ is parallel to $\hat p_{miss}$, and $\hat x$ and $\hat y$ are transverse to it and defined by: $\hat y\parallel \vec q \times \vec p_{miss}$ and $\hat x = \hat y\times \hat z$.

\begin{figure} [t]
\includegraphics[width=8cm,height=5.6cm]{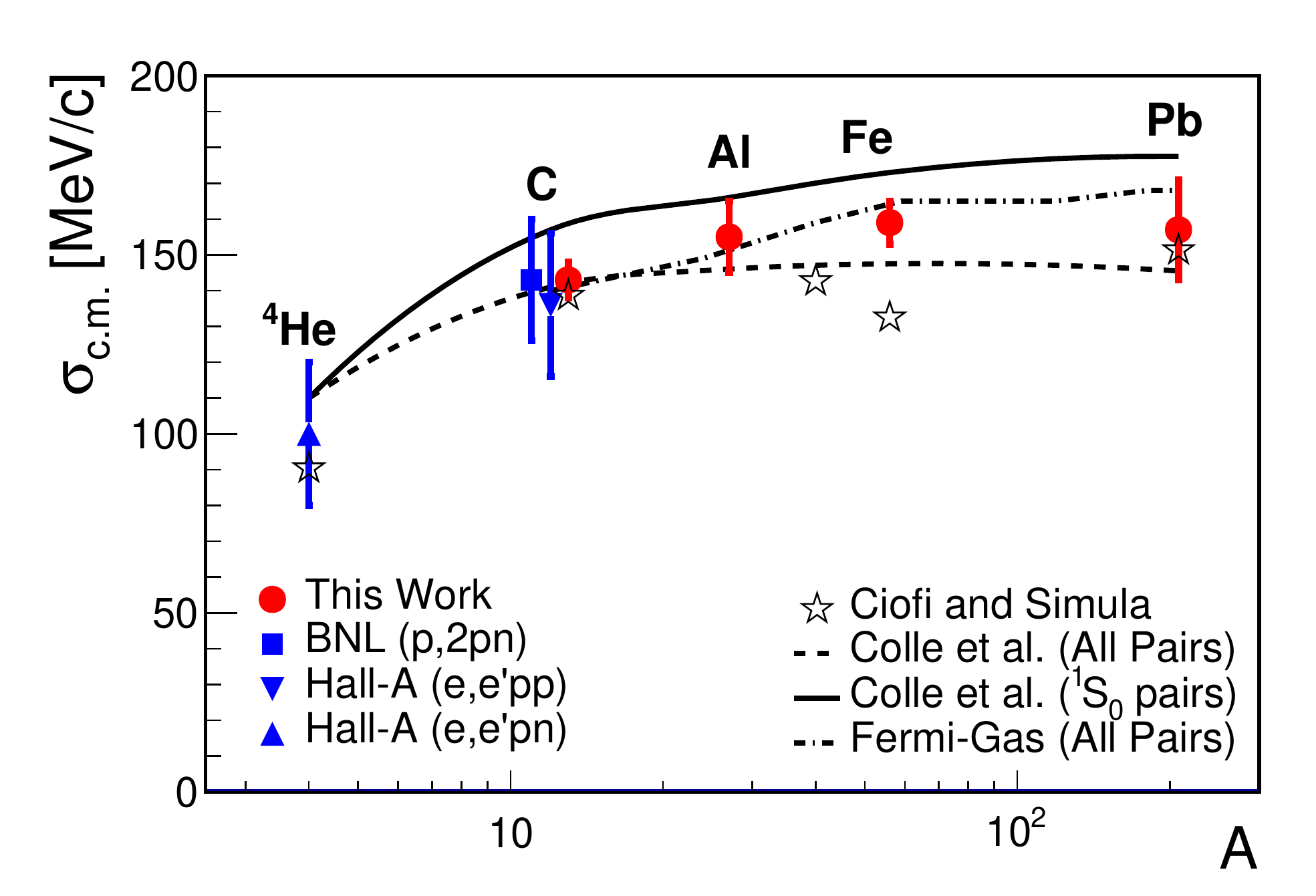}
\caption{\label{fig:2} (color online) The nuclear mass dependence of
  the one-dimensional width of the c.m. momentum distribution. The data points obtained in this work (red full circles) are compared to previous measurements (blue full squares and triangles)~\cite{tang03,shneor07,korover14} and theoretical calculations by Ciofi and Simula (open stars)~\cite{CiofidegliAtti:1995qe}, Colle et al., considering all mean-field nucleon pairs (dashed line) and only $^1S_0$ pairs (solid line)~\cite{Colle:2013nna} and a Fermi-gas prediction~\cite{moniz71} concidering all possible nucleon pairs. See text for details.}
\end{figure}

Figure~\ref{fig:1} shows the number of $A$\eepp{} events plotted versus the $x$- and $y$-
components of \pcm{} (see Eq.~\ref{eq:2}).  The data shown are not corrected for the CLAS
acceptance and resolution effects. As the $A$\eepp{} cross
section is proportional to $n^A_{c.m.}(\vec p_{c.m.})$, we can
extract the width of $n^A_{c.m.}(\vec p_{c.m.})$ from the widths of the
measured distributions.  Both $p^x_{c.m.}$ and $p^y_{c.m.}$
are observed to be normally distributed around zero for all nuclei.  
Thus, as expected, $n^A_{c.m.}(\vec p_{c.m.})$ can be approximated by a
three-dimensional Gaussian~\cite{tang03,shneor07,korover14,CiofidegliAtti:1995qe,Colle:2013nna},
and we characterize its width using $\sigma_x$ and $\sigma_y$, the
standard deviation of the Gaussian fits in the two directions
transverse to $\vec p_{miss}$. We average $\sigma_x$ and $\sigma_y$
for each nucleus to get $\sigma_{c.m.}$, the Gaussian width of one dimension of
$n^A_{c.m.}(\vec p_{c.m.})$.  These widths are independent of the
magnitude of $p_{miss}$, supporting the factorization of Eq.~\ref{eq:2}.

There are three main effects that complicate the interpretation of the raw (directly extracted) c.m. momentum distribution parameters (i.e., $\sigma_{c.m.}$): (1) kinematical offsets of the c.m. momentum in the $\hat{p}_{miss}$ direction, (2) reaction mechanism effects, and (3) detector acceptance and resolution effects. We next explain how each effect is accounted for in the data analysis.

(1) Kinematical offsets in the c.m. momentum direction: Since the relative momentum distribution of pairs falls rapidly for increasing $\vert\vec{p}_{rel}\vert$, it is more likely for an event with a large nucleon momentum ($\vec p_{miss}$) to be the result of a pair with smaller \prel{} and a \pcm{} oriented in the direction of the nucleon momentum. This kinematical effect will manifest as a shift in the mean of the c.m. momentum distribution in the $\hat{p}_{miss}$ (nucleon initial momentum) direction. 
To isolate this effect, we worked in a reference frame in which $\hat{z}\parallel \hat{p}_{miss}$ and $\hat x$ and $\hat y$ are perpendicular to $\hat p_{miss}$. The extracted c.m. momentum distributions in the $\hat x$ and $\hat y$ directions were observed to be independent of $\vec{p}_{miss}$, as expected.

(2) Reaction mechanism effects: These include mainly contributions from meson-exchange currents (MECs), isobar configurations (ICs),
and rescattering of the outgoing nucleons (final-state interactions or FSI) that can mimic the signature of SRC pair breakup and/or distort the measured distributions~\cite{groep00,Blomqvist:1998gq, Kester:1995zz}. 

This measurement was performed at an average $Q^2$ of about 2.1
GeV$^2$ and $x_B \geq 1.2$ to minimize the contribution of MEC and IC
relative to SRC
breakup~\cite{Arnold:1989qr,Laget:1987hb,Frankfurt:1996xx,Colle:2015lyl}. 
Nucleons leaving the nucleus can be effectively ``absorbed", where
they scatter inelastically or out of the phase-space of accepted
events. The probability of absorption ranges from about 0.5 for C to
0.8 for Pb \cite{hen12a,garino92,oneill95,abbott98,garrow02}. Nucleons
that rescatter by smaller amounts (i.e. do not scatter out of the phase-space of accepted
events) are still detected, but have their
momenta changed.  This rescattering includes both rescattering of the struck nucleon from its correlated
partner and from the other $A-2$ nucleons. Elastic rescattering of the struck nucleon from its correlated partner will change each of their momenta by equal and opposite amounts, but will not change \pcm{} (see Eq.~\ref{eq:2})~\cite{Frankfurt:1996xx,Colle:2015lyl}.
To minimize the effects of rescattering from the other $A-2$ nucleons, not leading to absorption, we selected largely anti-parallel kinematics, where $\vec{p}_{miss}$ has a large component antiparallel to $\vec{q}$~\cite{Frankfurt:1996xx}. Relativistic Glauber calculations show that, under these conditions, FSI are largely confined to within the nucleons of the pair~\cite{Frankfurt:1996xx,Sargsian:2001ax,Colle:2015lyl,frankfurt08b,Arrington:2011xs}.

The probability of the struck nucleon rescattering from the $A-2$
nucleons is expected to increase with $A$. Such rescattering, when not
leading to reduction of the measured flux (i.e., absorption), should
broaden the extracted c.m. momentum distribution. The measured widths
do not increase strongly with $A$.  This provides 
evidence that, in the kinematics of this measurement, FSI with the other
$A-2$ nucleons do not distort the shape of the measured c.m. momentum distribution,
in agreement with theoretical calculations~\cite{Frankfurt:1996xx,Sargsian:2001ax,Colle:2015lyl}.

In addition, Single Charge Exchange $(n,p)$ processes can lead to the detection of an $A$\eepp{} event that originate from the hard breakup of an $np$-SRC pair. While such SCX processes have relatively low cross-sections, the predominance of SRC pairs by $np$ pairs enhances its impact in measurements of the $A$\eepp{} reaction. Using the formalism of Ref.~\cite{Colle:2015lyl}, assuming the abundance of $np$-SRC pairs is $20$ times higher than that of $pp$-SRC pairs, we estimate that such SCX processes account for approximately 40\% of the measured $A$\eepp{} events. This is a large fraction that could impact the interpertation of the data. However, as $pp$- and $np$-SRC pairs are expected to have very similar c.m. momentum densities~\cite{Colle:2013nna, Colle:2015lyl}, this effect should not have a significant impact on the width of the c.m. momentum density.

(3) Detector acceptance and resolution effects: While CLAS has a large acceptance, it is not complete, and the measured c.m. momentum distributions need to be corrected for any detector related distortions. Following previous analyses~\cite{shneor07,subedi08,korover14}, we corrected for the CLAS acceptance in a 6-stage process: 
(1) We modeled the c.m. momentum distribution as a three-dimensional Gaussian, parametrized by a width and a mean in each direction. In the directions transverse to $\hat{p}_{miss}$ the widths were assumed to be constant and equal to each other ($\sigma_x=\sigma_y=\sigma_t$) and the means were fixed at zero. In the direction parallel to $\hat{p}_{miss}$, both the mean and the width were varied over a wide range. 
(2) For a given set of parameters characterizing the c.m. momentum distribution in step (1), we generated a synthetic sample of $A$\eepp\ events by performing multiple selections of a random event from the measured $A(e,e'p)$ events and a random $\vec{p}_{c.m.}$ from the 3D Gaussian. The combination of the two produced a sample of recoil protons with momentum ($\vec{p}_{recoil} = \vec{p}_{c.m.} - \vec{p}_{miss}$). 
(3) We determined the probability of detecting each recoil proton using GSIM, the GEANT3-based CLAS simulation~\cite{CLASGeant2}. 
(4) We analyzed the Monte Carlo events in the same way as the data to extract the c.m. momentum distributions and fit those distributions in the directions transverse to $\hat{p}_{miss}$ with a Gaussian to determine their reconstructed width. 
(5) We repeated steps (1) to (4) using different input parameters for the 3D Gaussian model used in step (1) and obtained a `reconstructed' $\sigma_t$ for each set of input parameters. $\sigma_t$ was varied between 0 and 300 MeV/$c$. The mean and width in the $\hat{p}_{miss}$ direction were sampled for each nucleus from a Gaussian distribution centered around the experimentally measured values with a nucleus dependent width (1$\sigma$) ranging from 45 to 125 MeV/$c$ for the mean and 30 to 90 MeV/$c$ for the width. The exact value of the width of the distribution is a function of the measurement uncertainty for each nucleus. It extends far beyond the expected effect of the CLAS acceptance.
(6) We examined the distribution of the generated vs. reconstructed
widths in the directions transverse to $\hat{p}_{miss}$ to determine
the impact of the CLAS acceptance on the measured values. 

The net effect of the acceptance corrections was to reduce the widths of the c.m. momentum distributions by 15--20 MeV/$c$ for each nucleus and to increase the uncertainties.

As a sensitivity study for the acceptance correction procedure, we
examined two additional variations to the event generator in the
$\hat{p}_{miss}$ direction: (A) a constant width of 70 MeV/$c$ and (B)
a width and mean that varied as a linear function of
$|{p}_{miss}|$. The variation among the results obtained using each
method was significantly smaller than the measurement uncertainties
and gives a systematic uncertainty of 7\%. We also performed a `closure' test where we input pseudo-data with known width and statistics that matched the measurements, passed it through the CLAS acceptance to see the variation in the `measured' width and then applied the acceptance correction to successfully retrieve the generated value.

The CLAS reconstruction resolution, $\sigma_{res}$, for the c.m. momentum of $pp$ pairs was measured using the exclusive $d(e,e' \pi^{-}pp)$ reaction and was found to equal 20 MeV/$c$. We subtracted this in quadrature from the measured c.m. width: $\sigma_{corrected}^{2} = \sigma_{measured}^{2} - \sigma_{res}^{2}$, which amounts to a small, 2--3 MeV/$c$, correction.

Figure~\ref{fig:2} shows the extracted $\sigma_{c.m.}=\sigma_t$, in the directions transverse to $\hat{p}_{miss}$, including acceptance corrections and subtraction of the CLAS resolution. The uncertainty includes both statistical uncertainties as well as systematical uncertainties due to the acceptance correction procedure.

The extracted value of $\sigma_{c.m.}$ for C is consistent with
previous C$(e,e'pp)$ measurements of
$\sigma_{c.m.}^{pp}$~\cite{shneor07} and C$(p,ppn)$ measurements of
$\sigma_{c.m.}^{pn}$~\cite{tang03}, with significantly reduced
uncertainty.
The extracted width
grows very little from C to Pb, and is consistent with a constant
value within uncertainties (i.e., it saturates). 
The saturation of $\sigma_{c.m.}$
with $A$ supports the claim that, in the chosen kinematics, FSI with the
$A-2$ nucleons primarily reduces the measured flux,
while not significantly distorting the shape of the
extracted c.m. momenutm distribution.

Figure~\ref{fig:2} also compares the data to several theoretical
predictions for the c.m. momentum of the nucleons which couple
to create the SRC pairs. Ref.~\cite{CiofidegliAtti:1995qe} considers
all possible $NN$ pairs  from shell-model orbits, while
Ref.~\cite{Colle:2013nna} considers both all pairs, and nucleons in a relative $^1S_0$ state (i.e., nodeless $s$-wave with
spin 0)~\cite{Vanhalst:2011es,vanhalst12}. The simplistic Fermi-Gas
prediction samples two random nucleons from a Fermi sea
with $k_F$ from~\cite{moniz71}.

The agreement of the data with calculations supports the theoretical
picture of SRC pair formation from temporal fluctuations of
mean-field nucleons~\cite{Hen:2016kwk}. The experimentally
extracted widths are consistent with the Fermi-Gas prediction and
are higher than the full mean-field calculations that
consider formation from all possible pairs. The data are lower
than the $^1S_0$ calculation that assumes restrictive
conditions on the mean-field nucleons that form SRC
pairs~\cite{Colle:2013nna}.

We note that the SRC-pair c.m. momentum distributions extracted from experiment differ from those extracted directly from ab-initio calculations of the two-nucleon momentum distribution. The latter are formed by summing over {\it all} two-nucleon combinations in the nucleus and therefore include contributions from non-SRC pairs. See discussion in Ref.~\cite{Weiss:2016obx}.

In conclusion, we report the extraction of the width of the
c.m. momentum distribution, $\sigma_{c.m.}$, for $pp$-SRC pairs from
$A$\eepp{} measurements in C, Al, Fe, and Pb. The new data is
consistent with previous measurements of the width of the
c.m. momentum distribution for both $pp$ and $pn$ pairs in
C. $\sigma_{c.m.}$ increases very slowly and might even saturate from
C to Pb, supporting the claim that final state interactions are
negligible between the two outgoing nucleons and the residual $A-2$
nucleus. The comparison with theoretical models supports the claim
that SRC pairs are formed from mean-field pairs in specific quantum
states. However, improved measurements and calculations are required
to determine the exact states.


\begin{acknowledgments}
We acknowledge the efforts of the staff of the Accelerator and Physics Divisions at Jefferson Lab that made this experiment possible. We are also grateful for many fruitful discussions with L.L. Frankfurt, M. Strikman, J. Ryckebusch, W. Cosyn, M. Sargsyan, and C. Ciofi degli Atti. The analysis presented here was carried out as part of the Jefferson Lab Hall B Data-Mining project supported by the U.S. Department of Energy (DOE). The research was supported also by the National Science Foundation, the Israel Science Foundation, the Chilean Comisión Nacional de Investigación Científica y Tecnológica, the French Centre National de la Recherche Scientifique and Commissariat a l’Energie Atomique the French-American Cultural Exchange, the Italian Istituto Nazionale di Fisica Nucleare, the National Research Foundation of Korea, and the UK’s Science and Technology Facilities Council. Jefferson Science Associates operates the Thomas Jefferson National Accelerator Facility for the DOE, Office of Science, Office of Nuclear Physics under contract DE-AC05-06OR23177. The raw data from this experiment are archived in Jefferson Lab’s mass storage silo. E. O. Cohen would like to acknowledge the Azrieli Foundation. 
\end{acknowledgments}

\bibliography{ppCM_bib}

\end{document}